\documentclass[11pt,eqs]{article}
\usepackage{latexsym,amsmath}

\def\cleq{\setcounter{equation}{0}}
\title{
The T-dual symmetries of a bosonic string
\thanks{Work supported in part by
the Serbian Ministry of Education and Science, under contract No. 171031.}}
\author{Lj. Davidovi\'c \thanks{e-mail: ljubica@ipb.ac.rs}  and B. Sazdovi\'c
\thanks{e-mail: sazdovic@ipb.ac.rs}\\
{\it Institute of Physics, University of Belgrade}\\
{\it Pregrevica 118, 11080 Belgrade, Serbia}}

\begin{document}
\maketitle
\begin{abstract}
We 
investigate whether the symmetry transformations
of a bosonic string are connected by T-duality.
We start with a standard closed string theory.
We continue with a modified open string theory,
modified to preserve the symmetry transformations possessed by the closed string theory.
Because the string theory is conformally invariant
world sheet field theory,
in order to find the transformations which preserve the physics,
one has to demand the isomorphism between the
conformal field theories corresponding to the initial and the transformed field configurations.
We find the symmetry transformations corresponding to the similarity transformation of the energy-momentum tensor, and find that their generators are T-dual.
Particularly, we find that the general coordinate and local gauge transformations are T-dual,
so we conclude that
T-duality in addition to the well known 
exchanges, transforms symmetries of the initial and its T-dual  theory into each other. 

\end{abstract}

\section{Introduction}
\cleq

One of the most important notions in  theoretical physics is a symmetry.
What is a symmetry of the string theory is not yet clear mainly because the theory itself is not yet
formulated in a background independent way, which would enlighten its deeper
principles. However, it is believed that the symmetry does exist, and 
that it will 
lead to finding the
physically indistinguishable solutions to the string equations
of motion and choosing the
correct vacuum \cite{MEBO,EO1}.

A string theory revealed a T-duality, a symmetry which is a consequence of the 
strings extended nature. T-duality connects seemingly different string theories
by exchanging, for example the characteristics of the 
strings momentum and winding \cite{GPR,GR,BL,SJ}.
So, it should exchange the symmetries of string theories as well.
If that holds, one can say that symmetries of  string theories always appear in pairs.

In this paper, we will investigate symmetry of  the space-time in which the bosonic string moves,
using a world-sheet formulation \cite{MEBO}.
The formalism differs from the usual,
where the symmetry is a
transformation of the 
space-time fields which leave the classical action invariant.
This concept of a symmetry does not apply now, because
only the world-sheet values of the  space-time fields appear in the string action.
Still, a symmetry should be a change in the space-time fields which does not change the
physics. 
So, suppose one considers a string theory with some chosen space-time field configuration, and a string theory
with a modified space-time configuration.
How one determines if these two descriptions are physically equivalent?
The string theory is conformally invariant
world sheet field theory.
The physics is determined by the conformal field theory, corresponding to the field configuration
in question. 
The transformation on the fields will be a symmetry, if the corresponding
conformal field theories are isomorphic \cite{MEBO,EO1}.

If one is given a conformal  field theory, one will obtain a 
physically identical conformal field theory by  performing a 
similarity transformation on the operators of the initial conformal field theory
$$\hat O\rightarrow e^{-i\hat\Gamma}\hat O e^{i\hat\Gamma}.$$
This transformation does not change the 
algebraic properties, so the new theory will be physically the same as the initial theory.
However, the transformation will 
 in general make changes to the world-sheet
energy-momentum tensor.
If these changes can be interpreted as the changes in
the space-time fields, then the latter are the symmetry transformations of the target space.

This idea was introduced in \cite{MEBO,EO},
where this automorphism of the operator algebra was seen as an analogue of the change of variables in a partition function.
The problem of finding symmetries was reduced to the problem of finding the operator generating the symmetry transformation.
The first investigations were treating the string massless fields, but the method was then generalized to treat conformal deformations of conformal field theory \cite{OE,EO2}.

The classical analogue of the similarity transformation is a transformation of an variable of interest by a Poisson bracket between a generator and a variable.
In this paper, we will apply these transformations to the string sigma model
energy-momentum tensor.
We will start by considering  a small modification of
the space-time background  $\Pi_{\pm\mu\nu}\rightarrow \Pi_{\pm\mu\nu}+\delta\Pi_{\pm\mu\nu}$.
It will induce the transformation of the energy-momentum tensor.
We will demand that the new energy-momentum tensor satisfies the Virasoro algebra as well.
This means that the transformed theory is physically equivalent to the initial theory,
or that these   field transformations are the symmetry of space-time theory. 
In this way we will find a transformation of space-time fields corresponding to a similarity transformation and the generator of this symmetry. 

We will consider both closed and the open string theory. For the open string 
we will consider a modified action which in comparison with the closed string action has an additional surface term which enables the invariance of the complete action to the general coordinate transformations and the gauge transformations, which are the symmetries of the closed string. 
For the open string theory, the boundary conditions can be satisfied by choosing either  
the Neumann or the Dirichlet boundary condition for every coordinate direction.
If the choice is made, the modified action surface term is
given in terms of the corresponding Neumann and the Dirichlet gauge fields.
The closed string symmetries remain the symmetries of the open string theory taking the appropriate 
transformation of these gauge field.

If one includes T-duality into consideration, one can conclude that the
general coordinate transformations and the local gauge transformations
are not independent.
Comparing their generators, using the T-dual coordinate transformation laws, one concludes that they are 
T-dual. Therefore, the symmetries are T-dual and the complete generator of symmetries is
self-dual.

%%%%%%%%%%%%%%%%%%%%%%%%%%%%%%%%%%%%%%%%%%%%%%%%%%%%%%%%%%%%%%%%%%%%%%%%%%%%%%%%%%%%%
\section{The bosonic string essentials}
\cleq

The quantization of the bosonic string theory, describing the string moving in 
a background consisting of
a space-time metric $G_{\mu\nu}$, a Kalb-Ramond field $B_{\mu\nu}$ and a dilaton field $\Phi$,
lead to a conclusion that in order to have a conformal invariance on the quantum level
the energy-momentum tensor components $\hat{T}_\pm(\varphi)$, with $\varphi=(G_{\mu\nu},B_{\mu\nu},\Phi)$, have to obey  the Virasoro algebras \cite{EO1,CFMP,EG}
\begin{eqnarray}\label{eq:vir}
\Big{[}\hat{T}_\pm(\varphi(\sigma)),\hat{T}_\pm(\varphi(\bar\sigma))\Big{]}&=&i\hbar
\Big{[}\hat{T}_\pm(\varphi(\sigma))+\hat{T}_\pm(\varphi(\bar\sigma))\Big{]}
\delta^\prime(\sigma-\bar\sigma),
\nonumber\\
\Big{[}\hat{T}_\pm(\varphi(\sigma)),\hat{T}_\mp(\varphi(\bar\sigma))\Big{]}&=&0.
\end{eqnarray}
From these conditions follow the space-time equations of motion
which space-time fields $G_{\mu\nu},B_{\mu\nu},\Phi$ have to obey.
In order to obtain the symmetries of the space-time equations of motion, one does not need to find their
explicit form.
It is sufficient to consider the transformations which do not change 
the above relations.

%%%%%%%%%%%%%%%%%%%%%%%%%%%%%%%%%%%%%%%%%%%%%%%%%%%%%%%
\subsection{The conformal gauge}

The action which was quantized \cite{BBS}, for a constant dilaton field reads
\begin{equation}\label{eq:action0}
S[x] = \kappa \int_{\Sigma} d^2\xi\sqrt{-g}
\Big[\frac{1}{2}{g}^{\alpha\beta}G_{\mu\nu}(x)
+\frac{\epsilon^{\alpha\beta}}{\sqrt{-g}}B_{\mu\nu}(x)\Big]
\partial_{\alpha}x^{\mu}\partial_{\beta}x^{\nu},
\quad (\varepsilon^{01}=-1),
\end{equation}
where the integration goes over two-dimensional world-sheet
$\Sigma$
with coordinates $\xi^\alpha$ ($\xi^{0}=\tau,\ \xi^{1}=\sigma$). 
$g_{\alpha\beta}$ is intrinsic world-sheet metric and
$x^{\mu}(\xi),\ \mu=0,1,...,D-1$ are the coordinates of the
D-dimensional space-time
 and
$\kappa=\frac{1}{2\pi\alpha^\prime}$.

Taking a conformal gauge $g_{\alpha\beta}=e^{2F}\eta_{\alpha\beta}$, the action becomes
\begin{equation}\label{eq:action1}
S[x] = \kappa \int_{\Sigma} d^2\xi\ 
\partial_{+}x^{\mu}
\Pi_{+\mu\nu}(x)
\partial_{-}x^{\nu},
\end{equation}
with
\begin{eqnarray}\label{eq:pib}
\Pi_{\pm\mu\nu}(x)=
B_{\mu\nu}(x)\pm\frac{1}{2}G_{\mu\nu}(x),
\end{eqnarray}
given in terms of the 
 light-cone coordinates
$\xi^{\pm}=\frac{1}{2}(\tau\pm\sigma),$
$\partial_{\pm}=
\partial_{0}\pm\partial_{1}.$

The momentum corresponding to $x^{\mu}$ is
\begin{equation}
\pi_\mu =
 \frac{\partial{\mathcal{L}}}{\partial{\dot{x}^{\mu}}} =
\kappa G_{\mu\nu}(x)\dot{x}^{\nu}
-2\kappa B_{\mu\nu}(x)x^{\prime\nu},
\end{equation}
and therefore the canonical Hamiltonian 
for the theory (\ref{eq:action1}) equals
\begin{equation}
{\cal H}_{C}=\frac{1}{4\kappa}(G^{-1})^{\mu\nu}\Big[
j_{+\mu}j_{+\nu}
+j_{-\mu}j_{-\nu}
\Big],
\end{equation}
where the currents $j_{\pm\mu}$ are given by
\begin{equation}\label{eq:jdef}
j_{\pm\mu}=\pi_\mu+2\kappa\Pi_{\pm\mu\nu}(x)x^{\prime\nu}.
\end{equation}
One can  rewrite the Hamiltonian 
in terms of the
energy-momentum tensor components
\begin{equation}\label{eq:tpm}
T_\pm=\mp\frac{1}{4\kappa}(G^{-1})^{\mu\nu}j_{\pm\mu}j_{\pm\nu},
\end{equation}
as
\begin{equation}
{\cal H}_{C}=T_{-}-T_{+}.
\end{equation}

%%%%%%%%%%%%%%%%%%%%%%%%%%%%%%%%%%%%%%%%%%%%

\subsection{The gauge invariant approach}
Let us in this subsection consider a string theory without a gauge fixing.
If one takes
the following
parametrization of the world-sheet metric tensor $g_{\alpha\beta}$ \cite{SP}
\begin{equation}
g_{\alpha\beta}=e^{2F}\hat g_{\alpha\beta}=\frac{1}{2}e^{2F}\left(\begin{array}{cc}
-2h^{-}h^{+} & h^{-}+h^{+}\\
h^{-}+h^{+} & -2       
\end{array}\right),
\end{equation}
with $h^{-}>h^{+}$,
the action (\ref{eq:action0}) becomes
\begin{equation}\label{eq:actionh}
S = 2\kappa \int_{\Sigma}d^2\xi\sqrt{-\hat{g}}\ {\hat{\partial}}_{+}x^{\mu}
\Pi_{+\mu\nu}
{\hat{\partial}}_{-}x^{\nu},
\end{equation}
where $\Pi_{\pm\mu\nu}$ is defined by (\ref{eq:pib})
and the partial derivative is given by
\begin{equation}
\hat{\partial}_{\pm}=\frac{\sqrt{2}}{h^{-}-h^{+}}(\partial_{0}+h^{\mp}\partial_{1}).
\end{equation}

Varying the action over $x^\mu$, one obtains the equations of motion
\begin{equation}\label{eq:eqmot}
\hat\nabla_\pm\hat\partial_\mp x^\mu
+\Gamma^\mu_{\mp\nu\rho}\partial_\pm x^\nu\partial_\mp x^\rho=0,
\end{equation}
with $\hat\nabla_\pm$ being the covariant derivative \cite{SP}, defined by
\begin{equation}
\hat\nabla_\pm x_{n}=(\hat\partial_\pm+n\hat\omega_\pm)x_{n},\quad\omega_\pm=\mp\frac{\sqrt{2}h^{\mp\prime}}{h^{-}-h^{+}},
\end{equation}
where $x_{n}$ is a scalar, vector or tensor and $n$ is the sum of its world-sheet indices,
taking 1 for plus and -1 for minus. 
The generalized connection is defined by
\begin{equation}\label{eq:gen}
\Gamma^\mu_{\pm\nu\rho}=\Gamma^\mu_{\nu\rho}
\pm B^\mu_{\ \nu\rho},
\end{equation}
given in terms of the Christoffel symbol
$\Gamma^\mu_{\nu\rho}
=\frac{1}{2}(G^{-1})^{\mu\sigma}(\partial_\nu G_{\rho\sigma}+\partial_\rho G_{\sigma\nu}
-\partial_\sigma G_{\nu\rho})$
and  the field strength of the field $B_{\mu \nu}$,
$B^\mu_{\ \nu\rho}=(G^{-1})^{\mu\sigma}B_{\sigma\nu\rho}=(G^{-1})^{\mu\sigma}(\partial_\sigma B_{\nu\rho}+\partial_\nu B_{\rho\sigma}+\partial_\rho B_{\sigma\nu})$.

The momentum corresponding to $x^{\mu}$ is
\begin{equation}
\pi_\mu =
 \frac{\partial{\mathcal{L}}}{\partial{\dot{x}^{\mu}}} =
\frac{\kappa G_{\mu\nu}(x)}{
h^{-}-h^{+}}
 \Big{(}2\dot{x}^{\nu}+(h^{+}+h^{-})x^{\prime\nu}\Big{)}
-2\kappa B_{\mu\nu}(x)x^{\prime\nu}.
\end{equation}
One can extract $\dot{x}^{\mu}$
 from the last equation, to obtain
\begin{equation}
\dot{x}^{\mu} = \frac{(G^{-1})^{\mu\nu}}{2\kappa}\Big{(}
h^{-}(\pi_{\nu}+2\kappa\Pi_{-\nu\rho}x'^{\rho})-
h^{+}(\pi_{\nu}+2\kappa\Pi_{+\nu\rho}x'^{\rho})\Big{)}.
\end{equation}
Using the currents (\ref{eq:jdef}), the coordinate derivatives over world-sheet parameters become
\begin{equation}
\dot{x}^{\mu} = \frac{(G^{-1})^{\mu\nu}}{2\kappa}\Big{(}
h^{-}j_{-\nu}
-
h^{+}j_{+\nu}\Big{)},
\end{equation}
\begin{equation}
x'^{\mu} = \frac{(G^{-1})^{\mu\nu}}{2\kappa}\Big{(}j_{+\nu}-j_{-\nu}\Big{)}.
\end{equation}
The canonical Hamiltonian density $\mathcal{H}_{c}=\pi_{\mu}\dot{x}^{\mu}-\mathcal{L}$ is
\begin{equation}
\mathcal{H}_{c}=-h^{-}
T_{+}
-h^{+}
T_{-},
\end{equation}
with $T_\pm$ defined in (\ref{eq:tpm}).
For $h^\pm=\mp 1$, one returns to the conformal gauge.

%%%%%%%%%%%%%%%%%%%%%%%%%%%%%%%%%%%%%%%%%%%%%
\section{The symmetries of space-time}
\cleq

In this section, we will
search for the symmetries of the space-time in which the closed and the open strings propagate,
as well as for their generators. We will investigate 
the change of the world-sheet energy-momentum tensor caused by 
the change in the space-time fields. We will demand the transformed energy-momentum tensor
$T_\pm+\delta T_\pm$ still obeys the classical analogue of the Virasoro algebra (\ref{eq:vir}).
%%%%%%%%%%%%%%%%%%%%%%%%%%%%%%%%%%%%%%%%%%%

The energy-momentum tensor components $T_{\pm}$ satisfy 
two independent copies of Virasoro algebra.
To find a symmetry of the equations of motion,
one should conclude what kind of transformation of fields
$\varphi\rightarrow\varphi+\delta\varphi$, 
and consequently of the 
energy-momentum tensor
\begin{equation}
\hat{T}_\pm(\varphi+\delta\varphi)=\hat{T}_\pm(\varphi)+\delta\hat{T}_\pm(\varphi),
\quad
\delta\hat{T}_\pm(\varphi)={_{G}\hat{T}^{\mu\nu}_\pm}\delta G_{\mu\nu}
+{_{B}\hat{T}^{\mu\nu}_\pm}\delta B_{\mu\nu}+{_{\Phi}\hat{T}_\pm}\delta\Phi,
\end{equation}
conserves the Virasoro algebra. One need not know the explicit form of the space-time equations of motion  to
find its symmetry transformations. In order to have a conserved Virasoro algebra,
one should find transformations for which
the following conditions are fulfilled
\begin{eqnarray}\label{eq:jed}
&&\!\!\!\!\!\!\!\!\!\!\!\!\Big{[}\hat{T}_\pm(\varphi(\sigma)),\delta\hat{T}_\pm(\varphi(\bar\sigma))\Big{]}
+\Big{[}\delta\hat{T}_\pm(\varphi(\sigma)),\hat{T}_\pm(\varphi(\bar\sigma))\Big{]}
=i\hbar
\Big{[}\delta\hat{T}_\pm(\varphi(\sigma))+\delta\hat{T}_\pm(\varphi(\bar\sigma))\Big{]}
\delta^\prime(\sigma-\bar\sigma),
\nonumber\\
&&\!\!\!\!\!\!\!\!\!\!\!\!\Big{[}\delta\hat{T}_\pm(\varphi(\sigma)),\hat{T}_\mp(\varphi(\bar\sigma))\Big{]}
+\Big{[}\hat{T}_\pm(\varphi(\sigma)),\delta\hat{T}_\mp(\varphi(\bar\sigma))\Big{]}
=0.
\end{eqnarray}

%%%%%%%%%%%%%%%%%%%%%%%%%%%%%%%%%%%%%%%%%%%
It is known \cite{O} that a similarity transformation 
 applied to $\hat T_\pm$
$$
\hat T_\pm\rightarrow e^{-i\hat\Gamma}\hat T_\pm e^{i\hat\Gamma},
$$
 ensures the physical equivalence of the corresponding theories,
makes the change in $\hat T_\pm$, which corresponds to a change in the space-time
fields, without changing the physics. This kind of change in the space-time fields is therefore a symmetry transformation. The similarity transformation implies that the change of $\hat T_\pm$ is just
\begin{equation}\label{eq:dT}
\delta \hat T_\pm(\varphi)=-i\Big[\hat\Gamma,\hat T(\varphi)\Big].
\end{equation}
One can confirm that the last relation solves the conditions for the Virasoro algebra conservation
(\ref{eq:jed}).

In the subsequent sections, we will be interested in finding the change in the space-time fields,
which transform $T_{\pm}$ in a way which preserves the classical version of the Virasoro algebra.
We will search for a generator $\Gamma_\Lambda$ (where $\Lambda$ is some parameter)
such that its Poisson bracket with energy-momentum components $T_\pm$ produces  the variation
$\delta T_\pm=\{\Gamma,T_\pm\}$,
equal to the 
change of energy-momentum tensor caused by the 
variation of fields $\varphi\rightarrow\varphi+\delta\varphi$.
If such a generator exists, then the previous variation
is a symmetry transformation of the space-time.

%%%%%%%%%%%%%%%%%%%%%%%%%%%%%%%%%%%%%%%%%%%
\subsection{T-duality of the closed string symmetry generators}

Our goal in this and the subsequent sections is to find the generators of the general symmetry
transformations corresponding to the similarity transformation.
So, let us suppose the background fields undergo a small change in value $\Pi_{\pm\mu\nu}\rightarrow \Pi_{\pm\mu\nu}+\delta\Pi_{\pm\mu\nu}$. 
Let us find the generators of the symmetries $\Gamma$, for this transformation of the background fields.
The currents change by
\begin{equation}
\delta j_{\pm\mu}=2\kappa\delta\Pi_{\pm\mu\nu}(x)x^{\prime\nu},
\end{equation}
and therefore
\begin{equation}\label{eq:deltat}
\delta T_\pm=\frac{1}{2\kappa}\delta\Pi_{\pm\mu\nu}j_\pm^\mu j_\mp^\nu.
\end{equation}
Let us determine the algebra of the currents (\ref{eq:jdef}). Using the standard Poisson brackets between the coordinates and the momenta
\begin{equation}\label{eq:spb}
\{x^\mu(\sigma),\pi_\nu(\bar{\sigma})\}=\delta^\mu_\nu\delta(\sigma-\bar\sigma),
\end{equation}
 one obtains
\begin{eqnarray}
\{j_{\pm\mu}(\sigma),j_{\pm\nu}(\bar\sigma)\}&=&\pm2\kappa
\Gamma_{\mp\mu,\nu\rho}\,
x^{\prime\rho}(\sigma)\delta(\sigma-\bar\sigma)
\pm2\kappa G_{\mu\nu}(x(\sigma))\delta^\prime(\sigma-\bar\sigma),
\nonumber\\
\{j_{\pm\mu}(\sigma),j_{\mp\nu}(\bar\sigma)\}&=&
\pm2\kappa
\Gamma_{\mp\rho,\mu\nu}\,
x^{\prime\rho}(\sigma)\delta(\sigma-\bar\sigma),
\end{eqnarray}
where the generalized connection is defined by (\ref{eq:gen}).
Consequently, the Poisson brackets between $T_\pm$, defined by (\ref{eq:tpm}), and currents are
\begin{eqnarray}\label{eq:pbtj}
\{T_\pm(\sigma),j_{\pm\mu}(\bar\sigma)\}&=&\pm
\frac{1}{2\kappa}
\Gamma_{\mp\nu,\mu\rho}
j_{\pm}^{\nu}j^\rho_{\mp}\delta(\sigma-\bar\sigma)
-j_{\pm\mu}(\sigma)\delta^\prime(\sigma-\bar\sigma),
\nonumber\\
\{T_\pm(\sigma),j_{\mp\mu}(\bar\sigma)\}&=&\pm
\frac{1}{2\kappa}
\Gamma_{\mp\rho,\nu\mu}
j_\pm^\nu j^\rho_\mp\delta(\sigma-\bar\sigma).
\end{eqnarray}
Finally, we obtain the Virasoro algebra
\begin{eqnarray}
\{T_\pm(\sigma),T_\pm(\bar\sigma)\}&=&-\Big{[}T_\pm(\sigma)+T_\pm(\bar\sigma)\Big{]}\delta^\prime(\sigma-\bar\sigma),
\nonumber\\
\{T_\pm(\sigma),T_\mp(\bar\sigma)\}&=&0,
\end{eqnarray}
in agreement with the condition (\ref{eq:vir}).

We will suppose the generator of the symmetries in the following form\begin{eqnarray}\label{eq:gdef}
{\cal G}={\cal G}_{+}+{\cal G}_{-},
\quad
{\cal G}_\pm=\int d\sigma\, \Lambda^\mu_\pm\big(x(\sigma)\big) j_{\pm\mu}(\sigma).
\end{eqnarray}
Using (\ref{eq:pbtj}), one obtains the Poisson brackets between $T_\pm$ and the generators 
\begin{eqnarray}\label{eq:rel}
\{T_\pm(\sigma),{\cal G}_\pm(\bar\sigma)\}&=&\pm\frac{1}{2\kappa}\Big[
\Gamma^\mu_{\mp\rho\nu}
\Lambda^\rho_\pm
+\partial_\nu \Lambda^\mu_\pm
\Big]j_{\pm\mu}j^\nu_\mp ,
\nonumber\\
\{T_\pm(\sigma),{\cal G}_\mp(\bar\sigma)\}&=&\pm\frac{1}{2\kappa}\Big[
\Gamma^\nu_{\mp\mu\rho}
\Lambda^\rho_\mp
+\partial_\mu \Lambda^\nu_\mp
\Big]j^\mu_\pm j_{\mp\nu}.
\end{eqnarray}
If one defines the generalized covariant derivatives by
\begin{eqnarray}\label{eq:kovdef}
D_{\pm\mu}\Lambda^\nu&=&\partial_\mu \Lambda^\nu+\Gamma^\nu_{\pm\rho\mu}\Lambda^\rho=
D_\mu \Lambda^\nu\pm B^\nu_{\ \rho\mu}\Lambda^\rho,
\end{eqnarray}
one rewrites (\ref{eq:rel}) as
\begin{eqnarray}
\{T_\pm(\sigma),{\cal G}_\pm(\bar\sigma)\}&=&\pm\frac{1}{2\kappa}
\Big(D_{\mp\nu}\Lambda^\mu_\pm\Big)j_{\pm\mu} j^\nu_\mp ,
\nonumber\\
\{T_\pm(\sigma),{\cal G}_\mp(\bar\sigma)\}&=&\pm\frac{1}{2\kappa}
\Big(D_{\pm\mu}\Lambda^\nu_\mp\Big)
j^\mu_\pm j_{\mp\nu}.
\end{eqnarray}

We know that $T_\pm$ transforms as (\ref{eq:deltat}),
therefore we search for a generator ${\cal G}={\cal G}_{+}+{\cal G}_{-}$ such that
\begin{equation}
\delta T_\pm=\{{\cal G},T_\pm\}=\frac{1}{2\kappa}\delta\Pi_{\pm\mu\nu}j_\pm^\mu j_\mp^\nu,
\end{equation}
which implies
\begin{equation}
\delta\Pi_{\pm\mu\nu}=\mp\Big(D_{\mp\nu}\Lambda_{\pm\mu}+
D_{\pm\mu}\Lambda_{\mp\nu}\Big).
\end{equation}
Taking 
\begin{equation}\label{eq:defxi}
\Lambda_{\pm\mu}=
\xi_\mu\pm\Lambda_\mu,
\end{equation} one obtains
\begin{eqnarray}\label{eq:st}
\delta G_{\mu\nu}&=&-2(D_\mu\xi_\nu+D_\nu\xi_\mu),
\nonumber\\
\delta B_{\mu\nu}&=&D_\mu\Lambda_\nu-D_\nu\Lambda_\mu-2B_{\mu\nu}^{\ \ \rho}\xi_\rho.
\end{eqnarray}

Using the currents (\ref{eq:jdef}) and the gauge parameter (\ref{eq:defxi}), we rewrite the generator ${\cal G}$ 
 as
\begin{equation}
{\cal G}=\int d\sigma\Big[
2\xi^\mu\pi_\mu+2\kappa(2\xi^\mu B_{\mu\nu}+\Lambda^\mu G_{\mu\nu})x^{\prime\nu}
\Big].
\end{equation}
To simplify the last expression, one can
define another gauge parameter
\begin{equation}
\widetilde\Lambda_\nu=
2\xi^\mu B_{\mu\nu}+\Lambda^\mu G_{\mu\nu}=\Lambda_\nu-2B_{\nu\mu}\xi^\mu,
\end{equation}
so that 
\begin{equation}\label{eq:gltilde}
{\cal G}=2\int d\sigma\Big[
\xi^\mu\pi_\mu+\widetilde\Lambda_\mu \kappa x^{\prime\mu}
\Big].
\end{equation}
In terms of the new parameter the Kalb-Ramond field transforms as
\begin{eqnarray}
\delta B_{\mu\nu}&=&
D_\mu\widetilde\Lambda_\nu-D_\nu\widetilde\Lambda_\mu+
2\Big[
D_\mu\big(B_{\nu\rho}\xi^\rho\big)
-D_\nu\big(B_{\mu\rho}\xi^\rho\big)
-\xi^\rho B_{\rho\mu\nu}\Big],
\end{eqnarray}
and the generator (\ref{eq:gltilde}) is rewritten as
\begin{equation}\label{eq:gkomp}
{\cal G}={\cal G}_\xi+{\cal G}_{\widetilde\Lambda}.
\end{equation}

Therefore, the closed string described by (\ref{eq:actionh})
 is invariant under the general coordinate transformations
\begin{eqnarray}\label{eq:gct}
\delta_\xi G_{\mu\nu}&=&-2(D_\mu\xi_\nu+D_\nu\xi_\mu),
\nonumber\\
\delta_\xi B_{\mu\nu}&=&-2\xi^\rho B_{\rho\mu\nu}+2(\partial_\mu b_\nu-\partial_\nu b_\mu),
\quad b_\mu=B_{\mu\nu}\xi^\nu,
\end{eqnarray}
with $D_\mu\xi_\nu=\partial_\mu\xi_\nu-\Gamma^\rho_{\mu\nu}\xi_\rho,$
and the local gauge transformations
\begin{eqnarray}\label{eq:gauge}
&&\delta_\Lambda G_{\mu\nu}=0,
\nonumber\\
&&\delta_{\Lambda} B_{\mu\nu}=\partial_\mu\Lambda_\nu-\partial_\nu\Lambda_\mu,
\end{eqnarray}
where we omit tilde from $\Lambda_\mu$.

If one keeps in mind T-duality, the present form of the generator offers interesting conclusions.
T-duality connects physically equivalent string sigma models, and the connection between T-dual string backgrounds
and their variables is  simplest in the constant background case.
In that case, the well known T-duality relation \cite{T0,T01}
$$\pi_\mu\cong\kappa x^{\prime\mu}$$
stands. So, T-duality interchanges the
sigma derivative of the coordinates with the momenta.
The consequence of the above relation to the generator of  symmetry
(\ref{eq:gkomp}),
is that its constituents turn out to be T-dual as well
$${\cal G}_\xi\cong{\cal G}_{\widetilde\Lambda},$$
which makes the complete generator ${\cal G}$ self-dual.
Because of this, the local gauge transformations and the general coordinate transformations are T-dual also.
The same relation however does not hold in more complicated backgrounds,
when T-duality is performed along the nonisometry directions or in backgrounds
without the global shift symmetry. These backgrounds where
 discussed in \cite{DST,DST1,DST2} where the generalized T-dualization procedure, applicable along the arbitrary space-time direction was presented and elaborated.

%%%%%%%%%%%%%%%%%%%%%%%%%%%%%%%%%%%%%%%%%
\section{The open string and its symmetries}
\cleq
The open string described by the same action as the closed string is not invariant to the above symmetries. The change in the action caused by the
general coordinate transformations is
\begin{eqnarray}
&&\delta_\xi S=2\sqrt{2}\kappa\int d\tau \xi^\nu\Big{[}
-h^{-}\Pi_{+\nu\mu}\hat\partial_{-}x^\mu
+h^{+}\Pi_{-\nu\mu}\hat\partial_{+}x^\mu
\Big{]}\Big{|}_{\sigma=0}^{\sigma=\pi},
\end{eqnarray}
for the equation of motion (\ref{eq:eqmot})
and the change by the local gauge transformations is
\begin{eqnarray}
&&\delta_\Lambda S=2\kappa\int d\tau\Lambda_\mu\dot{x}^\mu\Big{|}_{\sigma=0}^{\sigma=\pi}.
\end{eqnarray}
The first expression can be rewritten as
$$\delta S=-2\int d\tau\xi^\mu\gamma^{(0)}_\mu\Big{|}_{\sigma=0}^{\sigma=\pi},$$
where
\begin{eqnarray}
&&\gamma^{(0)}_\mu=-\sqrt{2}\kappa
\Big{[}
-h^{-}\Pi_{+\nu\mu}\hat\partial_{-}x^\mu
+h^{+}\Pi_{-\nu\mu}\hat\partial_{+}x^\mu
\Big{]}
\nonumber\\
&&\qquad=
\frac{\kappa G_{\mu\nu}(x)}{
h^{-}-h^{+}
}\,
\Big{[}
(h^{-}+h^{+})\dot{x}^\nu+2h^{-}h^{+}x'^{\nu}
\Big{]}
+2\kappa B_{\mu\nu}(x)\dot{x}^{\nu}.
\end{eqnarray}
The boundary conditions of 
the open string are given in terms of this variable 
\begin{equation}\label{eq:bc}
\gamma^{(0)}_\mu\delta x^\mu\Big{|}_{\sigma=0}^{\sigma=\pi}=0.
\end{equation}
In Ref. \cite{Z} the way to gain invariance to the transformation (\ref{eq:gauge}) was offered,
and in Refs. \cite{BS,S} the open string action  invariant under both  (\ref{eq:gct}) and (\ref{eq:gauge}) was presented,
which in comparison with the standard action has an additional surface term
\begin{equation}\label{eq:dod}
S_{bon}=2\int d\tau\Big[
\kappa A_\mu(x)\dot{x}^\mu
-\bar{A}_\mu(x)(G^{-1})^{\mu\nu}\gamma^{(0)}_\nu
\Big]\Big|_{\sigma=0}^{\sigma=\pi}\,.
\end{equation}
This term makes the open string theory invariant under both general coordinate and local gauge transformations, if the introduced 
vector fields $A_\mu$ and $\bar{A}_\mu$ transform as
\begin{eqnarray}
&&\delta_\Lambda A_\mu=-\Lambda_\mu,
\nonumber\\
&&\delta_\xi \bar{A}_\mu=-\xi_\mu.
\end{eqnarray}

For each of the coordinates one can fulfill the boundary conditions (\ref{eq:bc}), by choosing either
the Neumann or the Dirichlet boundary condition. If we mark the coordinates with the Neumann condition by $x^{a},\,a=0,1,\cdots,p$ and the coordinates with the Dirichlet condition by $x^{i},\,i=p+1,\cdots,D-1$, the surface term (\ref{eq:dod}) reduces to
\begin{equation}\label{eq:doddelta}
S_{bon}=2\int d\tau
\Big[
\kappa A^{N}_{a}(x)\dot{x}^{a}
-A^{D}_{i}(x)(G^{-1})^{ij}\gamma^{(0)}_{j}
\Big]\Big|_{\sigma=0}^{\sigma=\pi}\,,
\end{equation}
where
$A^{N}_{a}$ and $A^{D}_{i}$ are $(p+1)$- and $(D-p-1)$-dimensional vector gauge fields, first living on the $Dp$-brane and the second orthogonal to $Dp$-brane.
The Neumann vector field is as usual coupled to the coordinate time parameter derivative and the Dirichlet vector field is coupled to variable $\gamma^{(0)}_\mu$ related to the boundary condition, depending on both world-sheet parameter derivatives of the coordinates.

%%%%%%%%%%%%%%%%%%%%%%%%%%%%%%%%%%%%%%%%%%%%%%
\subsection{Field strengths}

It is well known that in the bosonic string action,
the surface term can be rewritten in the form of  the 
Kalb-Ramond term. If all the boundary conditions are Neumann, then the action on the boundary
\begin{equation}
S_{bon}=2\kappa\int d\tau
A^{N}_\mu(x)\dot{x}^\mu
\Big|_{\sigma=0}^{\sigma=\pi}\,,
\end{equation}
can be rewritten as
\begin{equation}\label{eq:gdn}
S_{bon}=\kappa\int d^{2}\xi\,
{\cal F}^{N}_{\mu\nu}\,
\varepsilon^{\alpha\beta}\partial_\alpha x^\mu\partial_\beta x^\nu,
\end{equation}
with 
\begin{equation}
{\cal F}^{N}_{\mu\nu}=\partial_\mu A^{N}_\nu(x)
-\partial_\nu A^{N}_\mu(x).
\end{equation}

For the arbitrary choice  of the boundary conditions the action on the boundary is given by (\ref{eq:doddelta}).
Let us restrict our investigation to the following metric
\begin{equation}
G_{\mu\nu}=
\begin{bmatrix} 
G_{ab} & 0 \\
0 & G_{ij}
\end{bmatrix}.
\end{equation}
In that case, (\ref{eq:doddelta}) can be 
rewritten using (\ref{eq:bc}) as
\begin{equation}
S_{bon}=2\kappa\int d\tau
\Big[
{\cal A}^{(0)}_\mu(x)\dot{x}^\mu
+{\cal A}^{(1)}_\mu x^{\prime\mu}
\Big]\Big|_{\sigma=0}^{\sigma=\pi},
\end{equation}
with
\begin{eqnarray}
{\cal A}^{(0)}_{a}=A^{N}_{a},&&
{\cal A}^{(0)}_{i}=
-\frac{h^{-}+h^{+}}{h^{-}-h^{+}}A^{D}_{i}
+2(BG^{-1})_{i}^{\ j}A^{D}_{j},
\nonumber\\
{\cal A}^{(1)}_{a}=0,&&{\cal A}^{(1)}_{i}=-2\frac{h^{-}h^{+}}{h^{-}-h^{+}}A^{D}_{i}.
\end{eqnarray}
In terms of the field strengths, the surface term becomes
\begin{equation}\label{eq:gd}
S_{bon} = 2\kappa \int_{\Sigma}d^2\xi\sqrt{-\hat{g}}\ {\hat{\partial}}_{+}x^{\mu}
{\cal F}_{\mu\nu}
{\hat{\partial}}_{-}x^{\nu},
\end{equation}
$${\cal F}_{\mu\nu}={\cal F}^{(a)}_{\mu\nu}+\frac{1}{2}{\cal F}^{(s)}_{\mu\nu},$$
with 
\begin{eqnarray}\label{eq:FaFs}
&&{\cal F}^{(a)}_{a b}= \partial_a A_b^N - \partial_b A_a^N  \, , 
\nonumber\\
&&{\cal F}^{(a)}_{ij}=
-\frac{h^{-}+h^{+}}{h^{-}-h^{+}}(\partial_{i}A^{D}_{j}-\partial_{j}A^{D}_{i})
+2\partial_{i}((BG^{-1})_{j}^{\ k}A^{D}_{k})
-2\partial_{j}((BG^{-1})_{i}^{\ l}A^{D}_{l}),
\nonumber\\
&&{\cal F}^{(s)}_{ab}=0,
\nonumber\\
&&{\cal F}^{(s)}_{i j}=
-\frac{2}{h^{-}}\partial_{i} A^{(1)}_{j}
+\frac{2}{h^{+}}\partial_{j} A^{(1)}_{i}.
\end{eqnarray}
The above calculations are done for the open string moving in the constant background fields. If the
coordinate dependent background is assumed, then the field strengths ${\cal F}^{(s)}_{ij}$ and ${\cal F}^{(a)}_{ij}$ will have an additional terms, coming from
$\frac{h^{+}-h^{-}}{h^{+}h^{-}}A_{i}^{(1)}\Gamma^{i}_{\mp\nu\rho}$.

Comparing the boundary actions 
 (\ref{eq:gd}) with the action (\ref{eq:actionh}), we conclude that the addition of the surface term has changed the background fields by
\begin{eqnarray}
&&G_{\mu\nu}\rightarrow G_{\mu\nu}+{\cal F}^{(s)}_{\mu\nu}\equiv{\cal G}_{\mu\nu},
\nonumber\\
&&B_{\mu\nu}\rightarrow B_{\mu\nu}+{\cal F}^{(a)}_{\mu\nu}\equiv{\cal B}_{\mu\nu}.
\end{eqnarray}

%%%%%%%%%%%%%%%%%%%%%%%%%%%%%%%%%%%%%%%%%%%
\subsection{The symmetry generators for the open string}

The metric and the Kalb-Ramond field are changed in comparison with the closed string case to
$G_{\mu\nu}\rightarrow{\cal G}_{\mu\nu}=G_{\mu\nu}+{\cal F}^{(s)}_{\mu\nu}$ and
 $B_{\mu\nu}\rightarrow{\cal B}_{\mu\nu}= B_{\mu\nu}+{\cal F}^{(a)}_{\mu\nu}.$
Instead of the transformations (\ref{eq:st}),
the open string symmetry transformations,
for a theory with mixed boundary conditions are
\begin{eqnarray}
\delta {\cal G}_{\mu\nu}&=&-2({\cal D}_\mu\xi_\nu+{\cal D}_\nu\xi_\mu),
\nonumber\\
\delta {\cal B}_{\mu\nu}&=&{\cal D}_\mu\Lambda_\nu-{\cal D}_\nu\Lambda_\mu
-2{\cal B}_{\mu\nu}^{\ \ \rho}\xi_\rho,
\end{eqnarray}
where
$$
{\cal D}_\mu \Lambda^\nu=
\partial_\mu \Lambda^\nu+{ \Gamma}^\nu_{\rho\mu}({\cal G})\Lambda^\rho
$$
is a covariant derivative corresponding to the metric ${\cal G}_{\mu\nu}$.
By ${\cal B}_{\mu\ \nu}^{\ \rho}$ we marked  a field strength of the Kalb-Ramond field.
The field itself is changed, however the field strength of the additional part is zero and therefore
${\cal B}_{\mu\nu\rho}=B_{\mu\nu\rho}.$

The generator of the transformation is
\begin{equation}
{\cal G}=\int d\sigma\Big[
2\xi^\mu{\bf \pi}_\mu+2\kappa(2\xi^\mu {\cal B}_{\mu\nu}+\Lambda^\mu {\cal G}_{\mu\nu})x^{\prime\nu}
\Big],
\end{equation}
with
\begin{equation}
\pi_\mu =
\kappa {\cal G}_{\mu\nu}(x)\dot{x}^{\nu}
-2\kappa {\cal B}_{\mu\nu}(x)x^{\prime\nu}.
\end{equation}
Using the explicit form of the transformed metric, we obtain the connection
\begin{equation}
{\Gamma}_{\mu,\nu\rho}({\cal G})=\Gamma_{\mu,\nu\rho}-2\partial_\nu\partial_\rho A^{D}_\mu.
\end{equation}

If one choses only Neumann boundary conditions, the metric remains the same as ${\cal F}^{(s)}_{\mu\nu}=0$, so that only Kalb-Ramond field changes, but not its field strength.

%%%%%%%%%%%%%%%%%%%%%%%%%%%%%%%%%%%%%%%%%%%%

\section{Conclusion}
\cleq
We considered the general coordinate and the local gauge transformations of the bosonic string,
and showed that they are T-dual to each other.
We started with the bosonic string theory in a conformal gauge and in a gauge invariant form. One of the purposes of the latter was
to find the separation of the Hamiltonian into two energy-momentum tensors satisfying two copies of the Virasoro algebras.
These tensors are represented as products of currents, which are used for defining the generators of symmetries that were investigated.
The generators were defined as
the integrals over the spatial world-sheet parameter of  the weighted currents.

Given a form of the generator, we investigated how it effects the variables, whose transformation is defined
by a Poisson bracket between the generator of  symmetry and the corresponding variable.
We where interested in finding the explicit form of the change in the energy-momentum tensor caused by such a transformation. This form of transformation is used as a classical analogue of the quantum transformation, known to preserve the Virasoro algebra between energy-momentum tensor components.
So, both initial and the transformed energy-momentum tensor describe the same physics.
Because of that,
we were interested in finding the small variations of background fields such that the change in energy-momentum tensor they cause is exactly the considered transformation. In fact, if there exists the generator
such that the described equality is possible, then the obtained transformations of space-time fields are the
symmetry of the theory. 

We considered the standard closed string theory,
and a modified open string theory which in comparison to the standard open string theory
has an additional surface term. This term was chosen in such a way to cancel the
obstacle for the closed string symmetries to be the open string symmetries as well.
Introduction of the Neumann vector fields on the boundary is equivalent to the change of Kalb-Ramond field.
In the case considered here however,
the vector fields are not coupled only to the time derivative of the coordinates but also to the
functions which define the boundary conditions on the string endpoints. Therefore, 
both metric and Kalb-Ramond field are changed by the surface term.

We found the transformations of the background fields, which transform the energy-momentum tensor 
in such a way that the Virasoro algebra is preserved. It turned out, that the generators of the symmetries
(\ref{eq:gltilde})
can be separated in a way that one part of generator, linear in $x^{\prime\mu}$ represents the generator of the 
well known local gauge transformation of the Kalb-Ramond field
and the other part, proportional to $\pi_\mu$, generates the general coordinate transformations.
Since these quantities are T-dual $\kappa x^{\prime\mu}\cong\pi_\mu$,
we conclude that these symmetries are T-dual.
So, we showed that T-duality has an additional feature, that it interchanges the symmetries of a theory.
The generator of global coordinate transformations and local gauge transformations of initial theory is T-dual to the generator of local gauge transformations and the
general coordinate transformations of the T-dual theory.

\end{document}